**The internal energy expression of a long-range interaction complex system and its statistical physical properties**

Yanxiu Liu[1], Cheng Xu[1], Zhifu Huang[1,*], Bihong Lin[1] and Jincan Chen[2]

[1]College of Information Science and Engineering, Huaqiao University, Xiamen 361021, People's Republic of China

[2]Department of Physics, Xiamen University, Xiamen 361005, People's Republic of China

Considering the interactions of two arbitrary particles, we obtain an internal energy expression of the complex system having long-range interactions. Based on the postulate of "equal-probability principle" for all microstates, the probability distribution function of the system is derived, and consequently, some main statistical physical properties of the system are revealed. It is important to find that the properties of a long-range interaction system are very closely dependent on the interaction coefficient and particle number.

________________________________

*Email: zfhuang@hqu.edu.cn

# 1. Introduction

For a system composed of N particles, each particle possesses certain energy. When the energy due to the interactions among these particles is negligible compared with the total energy of the system, extensivity could be adapted [1][2]. Due to the additivity, extensive quantities are linear functions of the system size. Macroscopic systems with short-range interactions are extensive. However, when long-range interactions in a system have to be considered and the interaction energy is not negligible, we are forced to generically abandon the additivity. Indeed, non-additivity is common in the real world, ranging from atomics [3] to stellar clusters [4]-[7]. In recent years, intense researches on q-Gaussian distribution [8] have been carried out since Tsallis gave an expression of entropy based on nonaddivities [9][10]. On the other hand, Pressé et al. [11] concluded that for modeling nonexponential distributions such as power laws, nonextensivity should be expressed through the constraints. Latella et al. [12] proved that the thermodynamic formalism with long-range interactions shows the clear role played by nonadditivity. However, some necessary interpretations such as the physical meaning of q-index cannot be given very well. Thus, it is very necessary to continuously investigate the concrete properties of some complex systems.

In the present work, we consider the interactions of two arbitrary particles in a system and derive the internal energy expression of the system. Through this expression, we further deduce a distribution about quantum states, which is expounded to be similar to the q-Gaussian distribution.

## 2. Internal energy of a long-range interaction system

For a physical system composed of N identical distinguishable particles, when the particles in the system may be regarded as noninteracting, the total internal energy is equal to the sum of the energy $\varepsilon_i$ (i=1,2,...,N) of individual particle, i.e.,

$$U = \varepsilon_1 + \varepsilon_2 + \varepsilon_3 + ... \varepsilon_N. \tag{1}$$

By considering the sum of energy levels instead of the sum of particles, Eq. (1) is rewritten as

$$U = \sum_l a_l \varepsilon_l, \tag{2}$$

where $a_l$ denotes the number of particles with energy level $\varepsilon_l$, and

$$N = \sum_l a_l \tag{3}$$

is the total number of particles.

For a system having long-ranged interactions, the interaction energy of the system cannot be neglected. It is important how to calculate the interaction energy of the system. In order to solve this question, we borrow the concept of the ISING model [13] in this paper. The ISING model assumes that the interactions of two first-neighbor particles are proportional to the product of their respective spins, giving us a significant inspiration. Distinguished from ISING model, we suppose that the interaction energy of two arbitrary particles is proportional to the product of their respective energy. Hence, the interaction energy of the system can be expressed as

$$U_2 = \lambda(\varepsilon_1\varepsilon_2 + \varepsilon_1\varepsilon_3 + ....\varepsilon_1\varepsilon_N + \varepsilon_2\varepsilon_3 + \varepsilon_2\varepsilon_4 + ...\varepsilon_2\varepsilon_N + ...\varepsilon_{N-1}\varepsilon_N), \tag{4}$$

where $\lambda$ is an equivalent interaction coefficient, which is assumed to be the same for two arbitrary particles. When a system has only three particles, the internal energy of

the system is given by

$$U = \varepsilon_1 + \varepsilon_2 + \varepsilon_3 + \lambda(\varepsilon_1\varepsilon_2 + \varepsilon_1\varepsilon_3 + \varepsilon_2\varepsilon_3). \tag{5}$$

If we regard particles 1 and 2 as a whole, their overall energy is $\varepsilon_{1+2} = \varepsilon_1 + \varepsilon_2 + \lambda\varepsilon_1\varepsilon_2$, and consequently, the internal energy of the system is determined by

$$U = \varepsilon_{1+2} + \varepsilon_3 + \lambda\varepsilon_{1+2}\varepsilon_3 = \varepsilon_1 + \varepsilon_2 + \varepsilon_3 + \lambda(\varepsilon_1\varepsilon_2 + \varepsilon_1\varepsilon_3 + \varepsilon_2\varepsilon_3) + \lambda^2\varepsilon_1\varepsilon_2\varepsilon_3. \tag{6}$$

Using the similar method, we can derive the internal energy of the system having N interaction particles as

$$U = \sum_i \varepsilon_i + \lambda \sum_{i,j(i\neq j)} \varepsilon_i\varepsilon_j + \lambda^2 \sum_{i,j,k(i\neq j\neq k)} \varepsilon_i\varepsilon_j\varepsilon_k + ... \lambda^{N-1} \prod_i \varepsilon_i. \tag{7}$$

Note that Eq. (7) can be simplified as

$$\lambda U + 1 = \prod_i (1 + \lambda\varepsilon_i). \tag{8}$$

By substituting the sum of particles by the sum of energy levels, Eq. (8) can be expressed as

$$\ln(\lambda U + 1) = \sum_l a_l \ln(1 + \lambda\varepsilon_l) \tag{9}$$

or

$$U = \frac{\exp[\sum_l a_l \ln(1 + \lambda\varepsilon_l)] - 1}{\lambda}. \tag{10}$$

In the $\lambda \to 0$ limit, Eq. (10) is recovered as $U = \sum_l a_l \varepsilon_l$. Eq. (10) implies the fact that the long-range interaction energy is nonadditive.

## 3. Statistical properties of a long-range interaction system

Based on the postulate of "equal-probability principle" for all microstates of a system

$$\ln \Omega = N \ln N - \sum_l a_l \ln a_l + \sum_l a_l \ln \omega_l, \qquad (11)$$

where $\omega_l$ is degeneracy. Using the method of Lagrange's undetermined multipliers

$$\delta(\ln \Omega - \alpha N - \beta U) = 0, \qquad (12)$$

we obtain the most probable distribution of particles as

$$a_l = \omega_l e^{-\alpha}(1+\lambda \varepsilon_l)^{-\gamma}, \qquad (12)$$

where $\gamma = \beta(1+\lambda U)/\lambda$, which depends on the parameters $\beta$ and $\lambda$ and the internal energy U. If the partition function

$$Z = \sum_l \omega_l (1+\lambda \varepsilon_l)^{-\gamma} \qquad (13)$$

is introduced, we can derive the relation between the internal energy and the partition function as

$$\ln(\lambda U+1)^{(\lambda U+1)/\lambda} = -N\frac{\partial \ln Z}{\partial \beta}. \qquad (14)$$

Because U is a function of $\{a_l, \varepsilon_l\}$, we can obtain the following relation: $dQ = \sum_l \frac{\partial U}{\partial a_l} da_l$. Using Eq. (10), one can obtain

$$\beta \sum_l \frac{\partial U}{\partial a_l} da_l = d\{-\sum_l [a_l \ln \frac{a_l}{\omega_l}]\}. \qquad (15)$$

Note that $\beta$ is an integrating factor of dQ. According to the definition of entropy $dS = dQ/T$, $1/T$ is also an integrating factor of dQ. Thus, it may be obtained that $\beta = 1/kT$, which is consistent with the classical statistical theory. By using Eqs. (11) and (15), the entropy can be expressed as

$$S = -k \sum_l (a_l \ln \frac{a_l}{N\omega_l}) \tag{16}$$

and

$$S = Nk(\ln Z - \beta \frac{\partial \ln Z}{\partial \beta}). \tag{17}$$

It is easy to verify that the expression of the entropy with the partition function $Z$ is consistent with the Boltzmann form.

A problem of the broad interest for complex systems is to infer the mathematical form of a probability distribution [14]. By using Eq. (12), the probability distribution function of quantum states can be expressed as

$$p(\varepsilon_r) = (1 + \lambda \varepsilon_r)^{-\gamma} / Z, \tag{18}$$

where $\varepsilon_r$ is the energy of an arbitrary quantum state distinguished from $\varepsilon_l$ and the partition function can be written as $Z = \sum_r (1 + \lambda \varepsilon_r)^{-\gamma}$. Substituting Eq. (18) into Eq. (16), we verify that the Shannon entropy [15]

$$S = -Nk \sum_r p_r \ln p_r \tag{19}$$

is still true. This shows that it is unnecessary to introduce other forms of entropy for long-range interaction systems.

## 4. Discussion

Using Eq. (18), we can plot the curves of the probability distribution function $p(\lambda \varepsilon)$ of the system varying with $\lambda \varepsilon$ for differently given values of $\gamma$, as shown in Fig.1. Note that $p(\lambda \varepsilon)$ quickly decreases with the increase of $\lambda \varepsilon$. It is natural. For a given equivalent interaction coefficient, the higher the energy level $\varepsilon$, the smaller the probability distribution function. Using Eqs. (10) and (18), we can also plot the curves of the ratio of the internal energy without interaction to the total internal energy $\sum_l a_l \varepsilon_l / U$ varying with the particle number N, as shown in Fig.2. Note that when the particle number N is not large enough, the value of $\sum_l a_l \varepsilon_l / U$ is very close to 1. However, when the particle number N is large enough, the value of $\sum_l a_l \varepsilon_l / U$ quickly decreases with the increase of N and then approaches to zero. It means that when there are long-range interactions in a system, the interaction energy of the system dependents on the particle number.

It is very significant to note that the well-known power-law distribution can be directly derived from Eq. (18). As $\varepsilon_r$ is large enough, Eq. (18) can be written as $p(\varepsilon_r) \propto \varepsilon_r^{-\gamma}$, which is just a well-known power-law distribution. In addition, when $q = 1/\gamma + 1$ and $\beta_T = \lambda \gamma$ are chosen, Eq. (18) can be expressed as

$$p(\varepsilon) = [1-(1-q)\beta_T \varepsilon]^{1/(1-q)} / Z . \tag{20}$$

The form of Eq. (20) is similar to that of the q-exponential distribution obtained in Refs. [8][10]. In recent researches, the q-exponential distribution is acquiring the attention of many branches of science. For example, Caruso et al. [16] observed that the probability distribution of energy differences of subsequent earthquakes in the World Catalog and Northern California is well fitted by a q-Gaussian with q=1.75, which corresponds to $\gamma = 4/3$, and the probability distribution function is shown in Fig.1. Pickup et al. [17] obtained that the interaction parameter is directly related to the normalized Tsallis nonextensive entropy parameter q=5/3 in generalized spin glass relaxation, which corresponds to $\gamma = 3/2$, and the probability distribution function is also shown in Fig.1. In addition, Cai et al. [18] found that the PDF of the detrended electroencephalogram signals is well fitted by a q-Gaussian distribution. DeVoe [19] found a q-Gaussian distribution in a trapped ion interacting with a classical buffer gas, It was also found in turbulence experiments by Combe et al. [20] that a q-Gaussian distribution is fitted well in the probability density function of the displacement fluctuations. Li et al. [21] found that the diffusion constant may depend on q-index in nonlinear Fokker-Planck equation. Du [22] found that the $q$-exponential of a sum can be applied as the product of the $q$-exponential based on the probabilistically independent postulate employed in nonextensive statistical mechanics. Krasnov [23] also found $q$-exponential type velocity distribution of the ions in the optical superlattice. Chen et al. [24] presented a new method based on the $q$-exponential distribution to suppress the noise and increase the accuracy of feature extraction for ultrasonic Lamb wave signals. This shows that Eq. (18) may be very meaningful. The contact between Eq. (18) and the q-Gaussian distribution is worthy to be further investigated.

## 5. Conclusions

Based on the Boltzmann statistics, we have deduced an internal energy expression of the complex system having long-range interactions and explained its main statistical physical properties. It is found that the internal energy of a long-range interaction system is nonextensive and the probability distribution function of the system is a similar q-exponential distribution. It may make researchers review the problems of a long-range interaction system from another brand-new angle. Applications of the internal energy expression and nonexponential distribution will be helpful for the correct interpretation of experimental results in some complex dynamical systems including not only in physics but also in biology, economics, and earthquakes.

ACKNOWLEDGEMENTS

Authors would like to thank Dr. Congjie Ou for his helpful discussions. This work has been supported by the National Natural Science Foundation (No. 11305064), and the Fujian Provincial Natural Science Foundation (No. 2016J01021).

Figure captions:

FIG. 1. The logarithm of $p(\lambda\varepsilon)$ as a function of $\lambda\varepsilon$.

FIG. 2. The curves of the internal energy without interaction to the total internal energy $\sum_l a_l \varepsilon_l / U$ varying with the particle number N for a long-range interaction system at $T = 300K$.

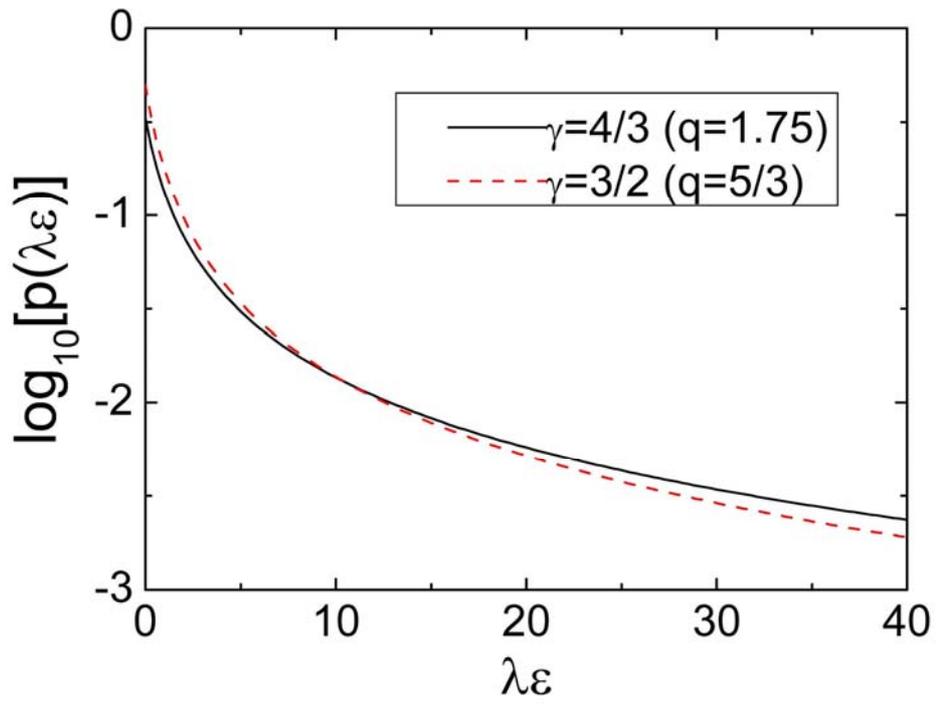

Fig. 1

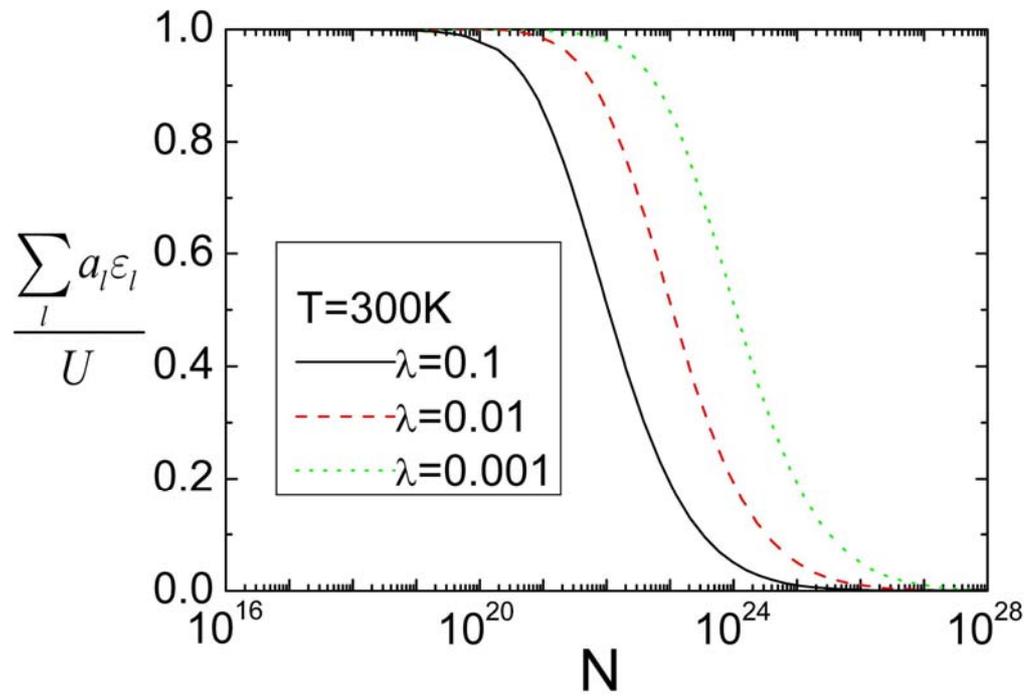

Fig. 2